\documentclass[journal=jctcce,manuscript=article]{achemso}
\setkeys{acs}{maxauthors=0,articletitle=true}
\usepackage{achemso}
\usepackage{placeins}
\usepackage{graphics}
\usepackage{amssymb,amsfonts}
\usepackage{graphicx}
\usepackage[table,dvipsnames]{xcolor}
\usepackage{multirow}
\usepackage{caption}
\usepackage{subcaption}
\usepackage{booktabs}
\usepackage{colortbl}
\usepackage{amsmath}
\usepackage{amsopn}
\usepackage{bm}
\usepackage{siunitx}
\usepackage{bm}
\usepackage{color}
\usepackage{array}
\usepackage{lscape}
\usepackage{mciteplus}
\usepackage[version=3]{mhchem}
\usepackage{ulem}
\usepackage{listings}
\usepackage{enumerate}
\usepackage{lmodern}

\DeclareMathOperator{\tr}{Tr}

\author{Frederik {\O}. Kjeldal}
\affiliation{DTU Chemistry, Technical University of Denmark\\Kemitorvet Bldg. 206, 2800 Kgs. Lyngby, Denmark}
\author{Janus J. Eriksen}
\email{janus@kemi.dtu.dk}
\affiliation{DTU Chemistry, Technical University of Denmark\\Kemitorvet Bldg. 206, 2800 Kgs. Lyngby, Denmark}

\title[TITLE]{Decomposing Chemical Space: Applications to the Machine Learning of Atomic Energies}

\begin{document}

\begin{abstract}
We apply a number of atomic decomposition schemes across the standard QM7 dataset---a small model set of organic molecules at equilibrium geometry---to inspect the possible emergence of trends among contributions to atomization energies from distinct elements embedded within molecules. Specifically, a recent decomposition scheme of ours based on spatially localized molecular orbitals is compared to alternatives that instead partition molecular energies on account of which nuclei individual atomic orbitals are centred on. We find these partitioning schemes to expose the composition of chemical compound space in very dissimilar ways in terms of the grouping, binning, and heterogeneity of discrete atomic contributions, e.g., those associated with hydrogens bonded to different heavy atoms. Furthermore, unphysical dependencies on the one-electron basis set are found for some, but not all of these schemes. The relevance and importance of these compositional factors for training tailored neural network models based on atomic energies are next assessed. We identify both limitations and possible advantages with respect to contemporary machine learning models and discuss the design of potential counterparts based on atoms and the intrinsic energies of these as the principal decomposition units.
\end{abstract}

\section{Introduction}

In standard formulations of electronic-structure theory, e.g., Kohn-Sham density functional theory (KS-DFT)~\cite{hohenberg1964inhomogeneous,kohn1965self,parr_yang_dft_book}, total properties are computed globally for chemical systems in their entirety on the basis of sets of nuclear coordinates and the finer details of the simulations, e.g., the one-electron basis set of atomic orbitals (AOs) and the exchange-correlation ($xc$) functional of choice. However, if one desires to decompose said simulations into local entities, for instance, amongst either the functional moieties, bonds, or atoms of a system at hand, this necessarily becomes a non-trivial task as there will be a strong element of ambiguity surrounding the infinitely many ways by which to partition a quantum-mechanical observable into the constituting parts of a quantum many-body system. Bar none, the most important principle any such partitioning scheme will need to satisfy is formal exactness in the decomposition of a chemical simulation (at a given level of theory), that is, the involved transformation from global to local properties must not be subject to any kind of loss. Other fundamental features include rapid and robust convergence of partitioned contributions upon an increase of the underlying AO basis set and a natural alignment of results with chemical intuition (with regards to electronegativities, molecular graphs, substituent effects, etc.).\\

In a recent study, we proposed a novel atomic partitioning scheme completely agnostic with respect to heuristics or supposed rules on the bond network of a chemical system~\cite{eriksen_decodense_jcp_2020}. Instead, the canonical molecular orbitals (MOs) of a standard mean-field simulation first undergo a unitary transformation into a spatially localized representation, and the contributions to total first-order properties, e.g., energies or dipole moments, from individual MOs next get weighted amongst the atoms of the molecule in question. While at least two degrees of freedom exist following this protocol, a set of optimal choices for both the localized MO basis and its atomic populations has been firmly established by now, effectively eliminating any arbitrariness ({\textit{vide infra}}). By avoiding explicit references to structural or bonding motifs, we have since shown how these decompositions can allow for probing local, atomic properties within condensed (bulk-like) phases~\cite{eriksen_local_condensed_phase_jpcl_2021}, but also the perturbations to a system's electronic structure that occur in optical transitions between different electronic states~\cite{eriksen_elec_ex_decomp_jcp_2022}.\\

The present study is concerned with the ability of decomposition schemes like our own to expose diversities among elements of the same type when these are embedded within different spatial environments across a small model subsection of chemical compound space. Besides comparing our own scheme to alternatives from the scientific literature on the basis of how systematic distributions of their results will be for different basis sets and density functional approximations (DFAs), we will also discuss the overall alignment of all results with chemical intuition. Finally, we are interested in assessing the accuracy by which atomic energies can be learned from tailored, custom neural network models trained on data from said partitioning schemes, but also the extent to which an augmentation of standard training sets by explicit atomic contributions will influence the prediction of total molecular energies.\\

Machine learning (ML), particularly when based on neural network (NN) architectures, is fast becoming a key component of quantum chemistry, given how data-driven models promise to allow for the treatment of much larger problem sizes than standard simulations formulated around electronic wave functions or densities~\cite{chmiela2018towards,unke2021machine,sauceda2022bigdml,qiao2020orbnet,christensen2021orbnet}. Arguably the most successful of such architectures is the high-dimensional variant (HDNN) of Behler and Parrinello~\cite{behler2007generalized}. In HDNNs, total properties of a molecule consisting of $\mathcal{M}$ atoms are obtained as a sum of intrinsic contributions associated with each of these. Either a single NN is used to predict all atomic contributions for a given system or separate NNs are implemented for each element. Predicted atomic properties are thus assumed to depend chiefly on local chemical environments, although corrections that account for long-range and dispersion effects may also be included~\cite{grimme_dft_d3_jcp_2010,grimme_disp_corr_mf_methods_chem_rev_2016}. There are multiple benefits to such a decomposition into atomic contributions since the prediction of total, molecular properties will scale as $\mathcal{O}(\mathcal{M})$ and the evaluation can be trivially parallelized. Furthermore, it is, at least in principle, possible to train models on small molecules before applying them to more extended problems, thereby allowing for some degree of transferability across chemical compound space. However, the vast majority of all such existing models have been trained exclusively on global data, e.g., molecular energies, which correlate with molecular composition through exceedingly intricate scaling relations.\\ 

Two main classes of HDNN models have emerged over the years; those based on descriptors and those formulated around message passing, differing from one another in how local chemical environments are represented~\cite{ceriotti_ml_representation_prl_2020,behler_lilienfeld_goedecker_ml_representation_mlst_2021}. Descriptor-based NNs operate in terms of fixed sets of such representations, encoded into so-called atomic environment vectors (AEVs) used as the input for a standard feed-forward NN~\cite{behler_ml_acsf_jcp_2011}. In contrast, message-passing models iteratively encode these local environments, whereby representations are granted the flexibility to automatically adapt to reference data by capturing more complex chemical interactions~\cite{tkatchenko_mueller_ml_qc_jcp_2018}.\\

As the production of quantum-chemical training data typically involves numerous, expensive simulations, it is beneficial to be able to extract the maximum amount of information from each of these, at the lowest possible overhead. To that end, the inclusion of nuclear forces for each atom as a training label allows for more fine-grained data in comparison to the case of training on total energies only. The extra cost of obtaining forces varies, but is generally of the same order of magnitude as the preceding energy calculation. The addition of gradient data has been demonstrated to improve upon the prediction of both forces and energies in applications to potential energy surfaces of single molecules, but when considering datasets whose entries vary in conformation and size, improvements in energy evaluations tend not to materialize~\cite{christensen2020role}. Also, the inclusion of forces is useful only whenever these do not vanish, i.e., for model applications to molecular structures distorted away from equilibrium.\\

We will here consider an alternative approach, namely, one in which atomic energies, rather than forces, are included as training labels, whereby existing datasets of total energies get augmented by an additional $\mathcal{M}$ data points for each entry (as opposed to thrice this number in the case of forces). Again, the overhead of performing a decomposition of total properties into their atomic constituents will depend on the exact scheme employed, but the cost may generally be assumed to be of the order of a single mean-field iteration. Such a strategy has previously been considered for solid-state applications, e.g., in the design of force fields~\cite{huang2019density, hu2020neural} and in the study of grain boundaries~\cite{song2022atomic}. In here, we will instead restrict our focus to chemical Hamiltonians only, seeking to evaluate the usefulness of training standard NNs on atomic, rather than solely total energies. For this purpose, we will decompose the standard QM7 dataset~\cite{blum2009970, rupp2012fast}, which consists of 7165 constitutional and structural molecular isomers each containing up to 7 heavy atoms (C, N, O, and S), with the largest molecule built from a mere total of 23 atoms. Despite its small extent, the dataset encompasses a number of fundamentally distinct local atomic environments. Sulfur atoms are the most scarce, nitrogens and oxygens exist in an approximately equal amount, while carbons and hydrogens are the most numerous, the latter being the far most common element in the dataset.\\

We will begin our study by comparing how various decomposition schemes distribute atomic energies for the entire QM7 suite, with a focus on {\textit{if}} and {\textit{how}} partitioned results align with expectation. We will discuss energies in terms of polarization and accepted trends in tendencies of electrons to be shared unevenly between atoms, but also any possible variances under a change of basis set or DFA. Next, we will proceed to study how both the aggregate and diversity in the representation of each element from these decompositions will influence the rate by which atomic energies may be learned using different kinds of NNs. Finally, we will end by discussing how these results can impact the ability to predict total energies as a sum of atomic contributions, touching upon the extent to which such constraints on NNs to yield specific contributions may inflict penalties on molecular property predictions, but also what opportunities such new local models may offer going forward, e.g., in relation to the prediction of properties associated with independent atoms, bonds, or functional moieties.

\section{Theory}

As outlined in Ref. \citenum{eriksen_decodense_jcp_2020}, a KS-DFT molecular energy may be partitioned amongst the atomic nuclei of a chemical system by recasting the energy functional into the following form
\begin{align}
E = \sum^{\mathcal{M}}_{K}E_{\text{elec},K}(\bm{D},\bm{\delta}_K) + E_{xc,K}(\bm{\rho},\bm{\varrho}_K) + E_{\text{nuc},K} \ . \label{atom_decomp_eq}
\end{align}
In Eq. \ref{atom_decomp_eq}, the nuclear and electronic contributions are defined as
\begin{subequations}
\label{atom_contr_eqs}
\begin{align}
E_{\text{nuc},K} &= \frac{Z_K}{2}\sum^{\mathcal{M}}_{K \neq L}\frac{Z_L}{|\bm{r}_K - \bm{r}_L|} \label{nuc_contr_atom_eq} \\
E_{\text{elec},K} &= \tr[\bm{T}_{\text{kin}}\bm{\delta}_K] + \tfrac{1}{2}(\tr[\bm{V}_{K}\bm{D}] + \tr[\bm{V}_{\text{nuc}}\bm{\delta}_{K}]) + \tfrac{1}{2}\sum_{\sigma}\tr[\bm{G}_{\sigma}(\bm{D})\bm{\delta}_{K,\sigma}] \label{elec_contr_atom_eq} \\
E_{xc,K} &= \tr[\epsilon_{xc}(\bm{\rho})\bm{\varrho}_K] \ . \label{xc_contr_atom_eq}
\end{align}
\end{subequations}
In Eq. \ref{nuc_contr_atom_eq}, $Z_K$ and $\bm{r}_K$ denote the nuclear charge and position of atom $K$, while the kinetic energy and nuclear attraction operators in Eq. \ref{elec_contr_atom_eq} are denoted by $\bm{T}_{\text{kin}}$ and $\bm{V}_{\text{nuc}}$, respectively, alongside the attractive potential associated with atom $K$, $\bm{V}_{K}$, and an effective Fock potential, $\bm{G}_{\sigma}$ ($\sigma = \alpha,\beta$ is an electronic spin index). In Eq. \ref{elec_contr_atom_eq}, $\bm{D}$ denotes the full, spin-summed 1-electron reduced density matrix (1-RDM), while the objects that principally define this decomposition---the atom-specific 1-RDMs, $\{\bm{\delta}\}$---are constructed as follows:
\begin{align}
\bm{\delta}_K &= \sum_{\sigma}\bm{\delta}_{K,\sigma} = \sum_{\sigma}\sum^{\mathcal{N}_{\sigma}}_{i}\bm{d}_{i,\sigma}\bm{p}^{K}_{i,\sigma} \ . \label{atom_rdm1_eq}
\end{align}
In turn, these are formulated via a set of 1-RDMs, $\bm{d}_{i,\sigma} = \bm{C}_{i,\sigma}\bm{C}^T_{i,\sigma}$, unique to the individual occupied spin-$\sigma$ MOs of the system, $\bm{C}_{i,\sigma}$, and a set of weights of all $\mathcal{N}_{\sigma}$ MOs of $\alpha$-/$\beta$-spin on a given atom $K$, $\{\bm{p}^{K}\}$. Our earlier investigations in Refs. \citenum{eriksen_decodense_jcp_2020,eriksen_local_condensed_phase_jpcl_2021,eriksen_elec_ex_decomp_jcp_2022} have convincingly indicated how the atomic weights used to assign $\{\bm{d}\}$ should ideally not be drawn from regular Mulliken population analyses~\cite{mulliken_population_jcp_1955}, but rather recast into a basis of reduced dimension. The $xc$ energy in Eq. \ref{xc_contr_atom_eq} is expressed in terms of the computed energy density, $\epsilon_{xc}$, as derived from the total electronic density, $\bm{\rho}$, and possibly its derivatives, which are all quantities that may be trivially defined in an atom-specific manner, $\{\bm{\varrho}\}$, by proceeding through the 1-RDMs, $\{\bm{\delta}\}$.\\

In most other alternative decomposition schemes, particularly those that are similarly exact (i.e., lossless), properties are instead partitioned amongst the constituent atoms on the basis of the full 1-RDM. In the arguably most intuitive example of such a decomposition, the energy density analysis (EDA) by Nakai~\cite{nakai_eda_partitioning_cpl_2002,nakai_eda_partitioning_ijqc_2009}, $\bm{D}$ is partitioned on account of which atoms the individual AOs are localized on (that is, irrespective of further population measures) by limiting trace operations in Eqs. \ref{atom_contr_eqs} to only those AOs that are spatially assigned to atom $K$. In here, we will compare the use of Eqs. \ref{atom_decomp_eq} and \ref{atom_contr_eqs} to the original EDA scheme. However, it should be noted how this scheme has since been extended by employing either natural atomic orbitals or real-space grids~\cite{baba2006natural,imamura2007grid}, both of which ameliorate the excessive basis-set dependence of the theory somewhat (albeit at the potential expense of numerical exactness in Ref. \citenum{imamura2007grid}).

\section{Computational Details}

For the QM7 dataset, {\texttt{PySCF}}~\cite{pyscf_wires_2018,pyscf_jcp_2020} was used to compute molecular energies using the PBE~\cite{perdew_burke_ernzerhof_pbe_functional_prl_1996}, B3LYP~\cite{becke_b3lyp_functional_jcp_1993,frisch_b3lyp_functional_jpc_1994}, and M06-2X~\cite{zhao_truhlar_m06_functional_tca_2008} $xc$ functionals in Jensen's segmented polarization-consistent basis sets~\cite{jensen_pc_basis_sets_jcp_2001} (pcseg-$x$, $x =$ 0--2), possibly augmented by appropriate diffuse functions (aug-). All decompositions were subsequently performed using the {\texttt{decodense}} package~\cite{decodense}, yielding atomization energies by subtracting gas-phase, free-atom energies from the decomposed results.\\

Both NN architectures based on fixed descriptors and message-passing techniques were investigated due to their different processing of local chemical environments. The acclaimed {\texttt{ANI}} model~\cite{smith2017ani}, as implemented in the {\texttt{TorchAni}} Python package~\cite{gao2020torchani}, was used to train an example of the former kind, whereas the {\texttt{PhysNet}} code was used to train a message-passing analogue~\cite{unke2019physnet}. Throughout Section \ref{results_section}, we will restrict our focus to results from the {\texttt{PhysNet}}-based model, since the standard {\texttt{ANI}} model was found to yield slightly worse results on the whole, while being insensitive to training on a combination of both atomic and molecular energies as opposed to training on the latter only (as is standard practice, cf. Sect. \ref{results_section}). Instead, these additional results have all been collected in the supporting information (SI).\\

We will collectively refer to the combined set of training and validation data as our training set, and the ratio between the two components was fixed to 4:1 throughout, with all remaining molecules constituting the accompanying test set~\bibnote{We used combined training and validation sets of 100, 500, 1000, 1433, 2866, 4299, and 5732 molecules, corresponding to 1.4, 7.0, 14.0, 20.0, 40.0, 60.0, and 80.0 $\%$ of the total QM7 dataset, respectively.}. Each model was trained five times, i.e., on sets of five seeds with $x$ random training and validation molecules, and default hyperparameters were used throughout (i.e., no cross-validations were performed)~\bibnote{The default hyperparameters from the {\texttt{TorchAni}} and {\texttt{PhysNet}} sources were used throughout. Please see Refs. \citenum{smith2017ani,gao2020torchani,unke2019physnet} for further details.}.

\section{Results and Discussion}\label{results_section}

\begin{figure}[ht!]
    \centering
    \includegraphics[width=0.85\textwidth]{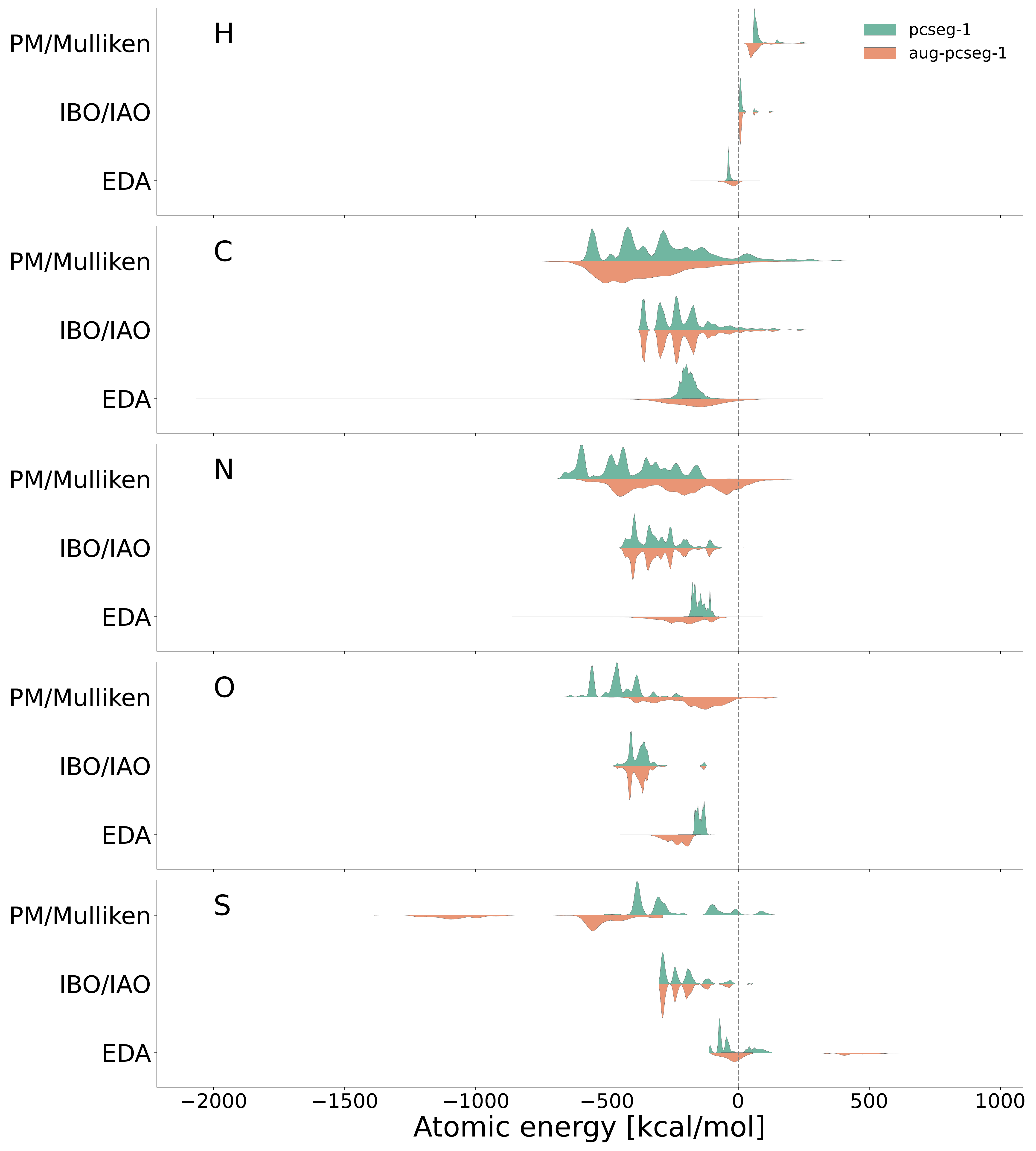}
    \caption{B3LYP/(aug-)pcseg-1 atomization energies for all elements in the QM7 dataset.}
    \label{fig_1}
\end{figure}
In Fig. \ref{fig_1}, we present distributions of atomization energies across all of QM7 for two different models formulated around Eqs. \ref{atom_decomp_eq} and \ref{atom_contr_eqs}---using a combination of either Pipek-Mezey~\cite{pipek_mezey_jcp_1989} (PM) localized MOs and standard Mulliken charges or intrinsic bond and atomic orbitals~\cite{knizia_iao_ibo_jctc_2013,knizia_visscher_iao_jctc_2021} (IBO/IAO)---alongside the original EDA scheme. As is obvious, these three schemes all yield vastly different results, both in the way they expose any underlying structure in the data, but also in their sometimes highly skewed responses to an augmentation of the double-$\zeta$ pcseg-1 basis set with additional diffuse functions. The PM/Mulliken distributions for the different elements are found to greatly overlap, as are the corresponding EDA distributions, while this trend is much less pronounced in the case of the IBO/IAO results, which appear much more structured in general, i.e., binned into distinct groups of contributions, than the other two. In the pcseg-1 basis set, the EDA results are observed to be characteristically uniform, spanning only a restricted energy interval across all different elements, while this is all but the case upon moving to the corresponding aug-pcseg-1 basis set. The same observation holds true for the PM/Mulliken results, as these also change dramatically upon adding more diffuse AOs. In the case of the IBO/IAO results, however, these are practically mirror images of one another in the two basis sets, which is satisfactory on its own, but also works to corroborate previous conclusions from Refs. \citenum{eriksen_decodense_jcp_2020,eriksen_local_condensed_phase_jpcl_2021,eriksen_elec_ex_decomp_jcp_2022} on the pronounced stability under a change of basis set of this particular decomposition scheme. In the SI, Fig. S1 presents corresponding atomization energies computed in corresponding single-$\zeta$ and triple-$\zeta$ basis sets (pcseg-0 and pcseg-2, respectively), and these results further support the observations drawn from Fig. \ref{fig_1}, namely, the stability of the IBO/IAO scheme and the striking lack of this property for the other two schemes. The corresponding variances of results under a change of DFA, on the other hand, are significantly less severe in the case of all three schemes, bar a near-constant shift in all atomization energies, cf. the results in Fig. S2 of the SI.\\

\begin{figure}[ht!]
    \centering
    \includegraphics[width=\textwidth]{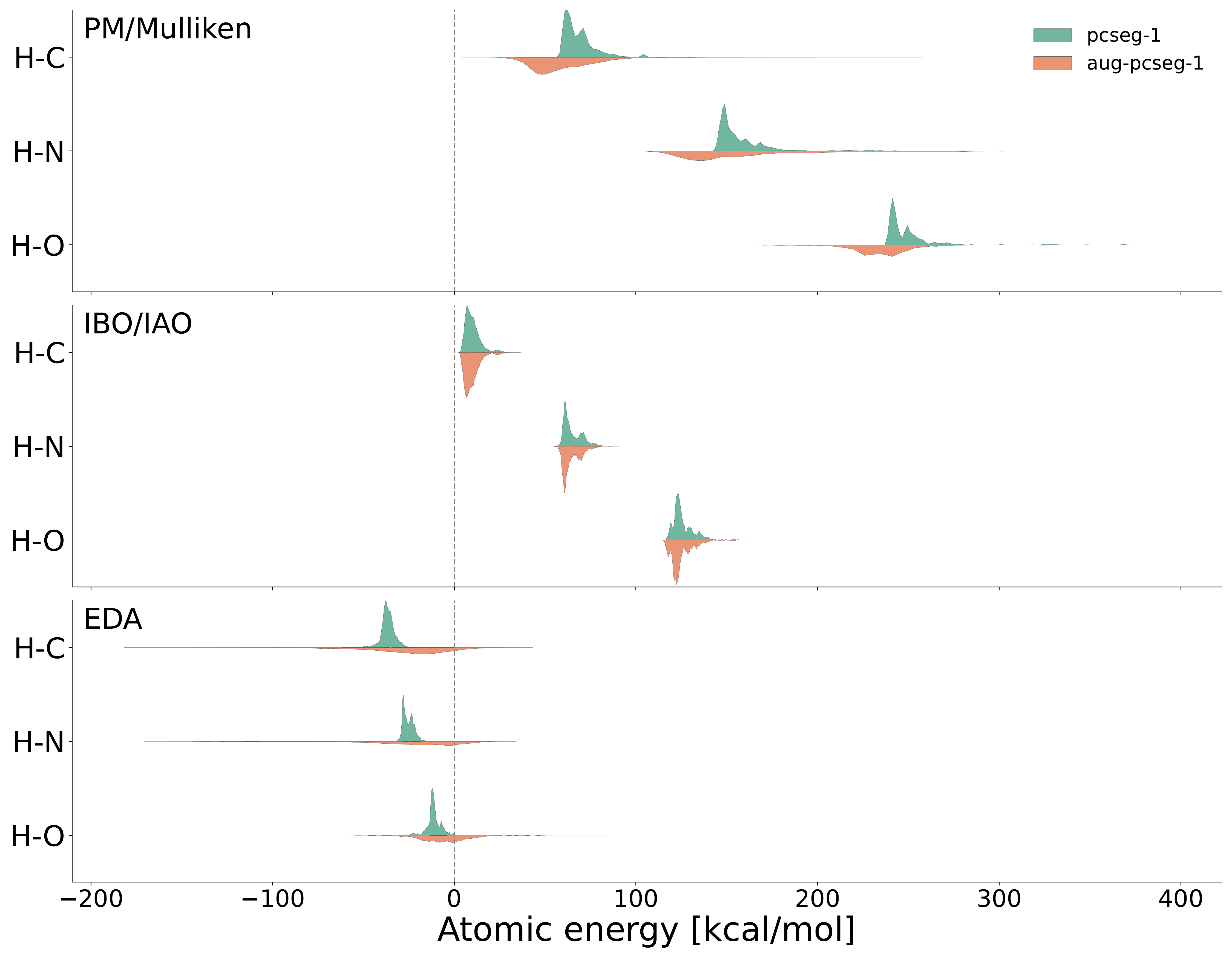}
    \caption{Subsection of Fig. \ref{fig_1} for all hydrogens in the QM7 dataset.}
    \label{fig_2}
\end{figure}
Next, in Fig. \ref{fig_2}, we zoom in on the hydrogen results from Fig. \ref{fig_1}, which may be grouped based on their neighbouring heavy atom (either C, N, or O). For this subset of hydrogen atoms, the atomization energies are all predicted to be negative in the EDA model in the pcseg-1 basis, and predominantly so in the aug-pcseg-1 basis, while all corresponding results of the two other schemes are positive. Recalling that positive and negative results in this frame imply destabilization and stabilization of an atom with respect to its resting state in vacuum, respectively, these variances indicate major differences in how the individual schemes represent the local electronic structures around the hydrogens in question. In the specific case of a hydrogen atom, it is arguably fair to assume results to be correlated with the electronegativity of its bonded neighbour. In Fig. \ref{fig_2}, the MO-based models of Ref. \citenum{eriksen_decodense_jcp_2020} predict exactly this, whereas the EDA scheme generally predicts a lowering of the energy of a hydrogen atom upon its embedding in a molecule. A clear separation between distinct types of hydrogens is observed for the IBO/IAO scheme, in particular, as these energies are clearly binned for the three classes. For the EDA partitioning, on the other hand, it appears as if results are principally determined by (while scaling with) the composition and extent of the underlying basis set due to the dependence on the full 1-RDM. The EDA results in the aug-pcseg-1 basis, in particular, are testament to this, unlike the decompositions of Ref. \citenum{eriksen_decodense_jcp_2020}, which allow for distinguishing different, unique atomic environments in any basis set.\\

These observations are further supported by the results in Fig. S3 of the SI, in which atomization energies of individual oxygen atoms are grouped by some of the numerous functional moieties that contain these. In here, the PM/Mulliken energies span the largest range, followed by the IBO/IAO and EDA energies. As for the hydrogen energies in Fig. \ref{fig_2}, the EDA distributions in Fig. S3 are once again observed to be notably uniform in the pcseg-1 basis, while any such structure in the data appears to vanish upon adding diffuse functions. For all three schemes, results for oxygens bonded to hydrogen and either of carbon or nitrogen have little overlap in the smaller of the two basis sets (true only for the IBO/IAO results in both of these), but how well these align with chemical intuition is seen to differ. The oxygen atoms in sulfone groups clearly have very peculiar energies compared to all other oxygens in both of the MO-based models of Ref. \citenum{eriksen_decodense_jcp_2020}, while the EDA scheme yields energies largely on par with the other oxygens. The exact reasons as to why local electronic structures around oxygens bonded to sulfur are so different, resulting in a characteristically singular shift upwards towards less stabilization in moving from, say, carbonyl to sulfone oxygens, is clearly worthy of further investigation, but unfortunately fall outside the scope of the present work.\\

\begin{figure}[ht!]
    \centering
    \includegraphics[width=\textwidth]{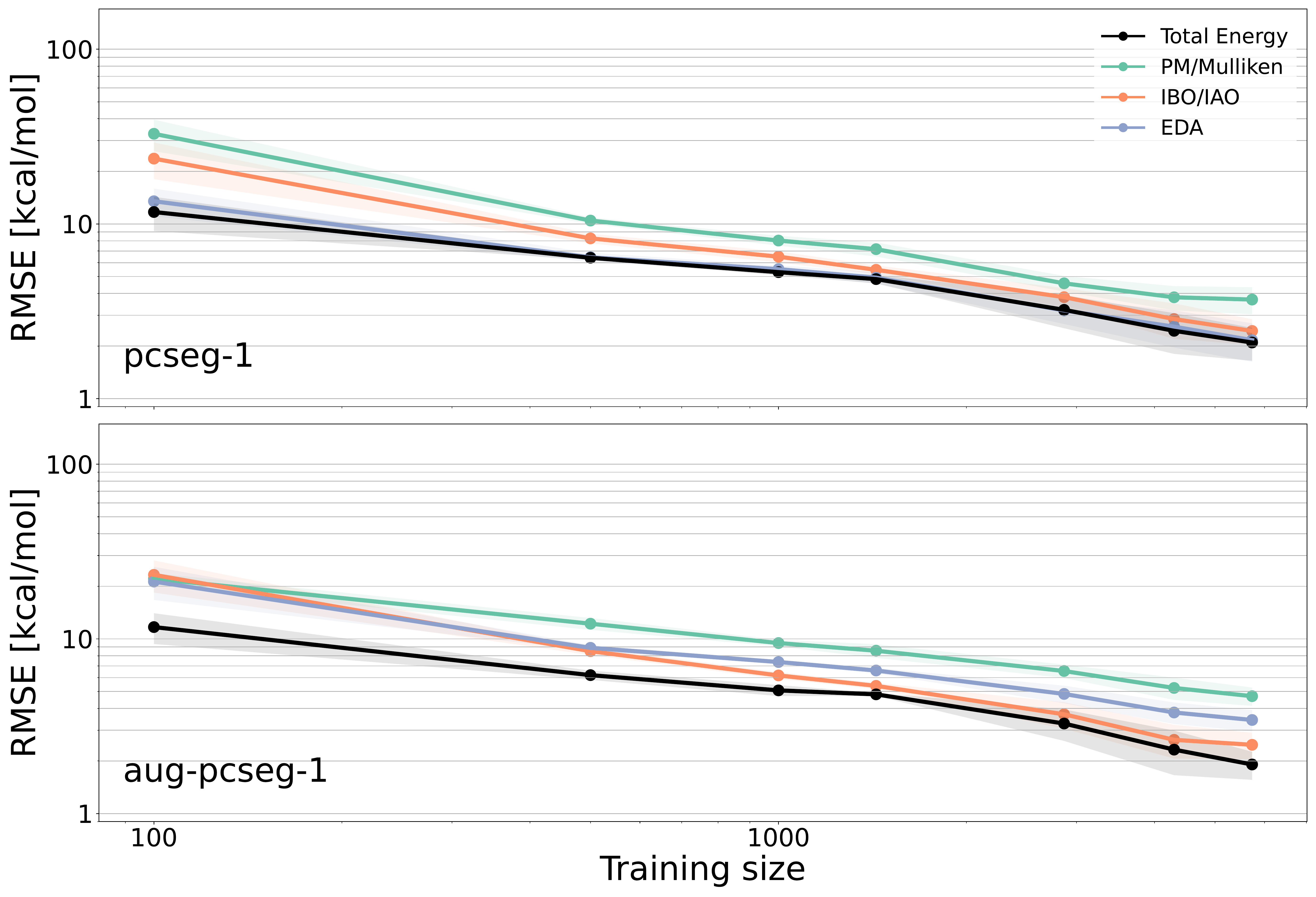}
    \caption{Root-mean-square error (RMSE) of predicted total energies for the QM7 dataset. Results are reported for NN models trained on total energies, but also NNs trained on each set of atomic energies from Fig. \ref{fig_1}. Shaded areas denote standard deviations of the RMSE.}
    \label{fig_3}
\end{figure}

Now, how well can we learn to predict atomic energies from each of the three models discussed so far, and how accurate are the total energies derived from these in comparison with standard NN implementations that only learn total, molecular energies? Comparing the total energy training curves in Fig. \ref{fig_3}, it is observed that standard training on total energies yields lower root-mean-square errors (RMSEs) than training on a combination of molecular and atomic energies from Fig. \ref{fig_1}, albeit only for limited training sets. Upon increasing the amount of data available to the models, their performances align with one another. For smaller training sets, the best atomic partitioning scheme is observed to be EDA, but only so in the pcseg-1 basis set. When training on EDA/aug-pcseg-1 data instead, the NN model is notably less capable of learning the underlying physics. The use of IBO/IAO data, on the other hand, is seen to result in somewhat larger errors for sparse training sets, but this NN model clearly learns more rapidly upon enlarging the combined training and validation set, yielding similar RMSE values as the model based on EDA data for the largest sets (and with identical results in both basis sets). The worst partitioning scheme for this purpose appears to be PM/Mulliken, with derived models yielding the largest RMSE values for all datasets, although the learning rate is still better in the pcseg-1 basis set than the EDA-based model.\\

All of these observations are further supported by the results in Fig. S4 of the SI, in which training curves similar to those in Fig. \ref{fig_3} are presented for {\texttt{PhysNet}}-based models trained exclusively on atomic energies. Here, the model trained on EDA data in the augmented basis set performs as poorly as that trained on PM/Mulliken, learning hardly anything at all upon being offered more training data. From Fig. S4, it is also worth noting how the RMSEs of all models exceed those reported in Fig. \ref{fig_3}, a point which is further substantiated in Fig. S5 where training curves are compared for {\texttt{PhysNet}}-based models that weigh (emphasize) atomic and molecular energies differently. While errors in predictions of individual atomic contributions are approximately alike in-between models, the availability of total energies still manages to regularize atomic contributions to such an extent that they accumulate to yield fair molecular energies (cf. Fig. \ref{fig_3}). For models based on fixed descriptors, on the other hand, no such improvements are observed from augmenting traditional datasets of total energies by additional atomic counterparts, cf. the {\texttt{ANI}}-based results in Figs. S12--S14 of the SI. We ascribe this insensitivity to the fact that the representations associated with individual atomic environments cannot relax in these models (by design), unlike more flexible, message-passing analogues like the custom {\texttt{PhysNet}}-based models of the present study.\\

The decline in performance of the models trained on a combination of atomic and total energies, rather than just the latter, for limited training sets can be traced back to two main attributes, namely, a lack of layers in the training data and an increase in the scarcity of the data. First, by design, the standard {\texttt{PhysNet}} and {\texttt{ANI}} models are free to vary their predicted atomic energies, and optimal values of these can thus be readily obtained for any given set of training data. How and whether these align with chemical intuition is somewhat irrelevant in this context, given that only their accumulated sum is evaluated. Furthermore, both types of NN models may be thought of as performing an atomic decomposition of molecular data themselves as, upon training, an implicit type of linear fit of all total energies as a function of the different elements provides a set of reference values for these, the sum of which gets subtracted from the property to be learned (typically accounting for more than $95\%$ this). The so-called self-interaction atomic energies (SIAEs) of the standard {\texttt{ANI}} model are prime examples of these references. Such a procedure will inevitably lead to a decrease in the standard deviation across a given training set, resulting in total quantities to be predicted that fall within a much more reasonable numerical range, thereby lowering the RMSE. In contrast, models trained on atomic energies are constrained to learn specific values of these from a partitioning scheme of choice, and these contributions hence cannot be decomposed further through the introduction of an intermediate, SIAE-like basis. Be that as it may, Fig. \ref{fig_3} still demonstrates how total energies---derived as a sum of atomic contributions---can be successfully learned whenever NN models are trained on physically sound datasets.\\

\begin{figure}[ht!]
    \centering
    \includegraphics[width=\textwidth]{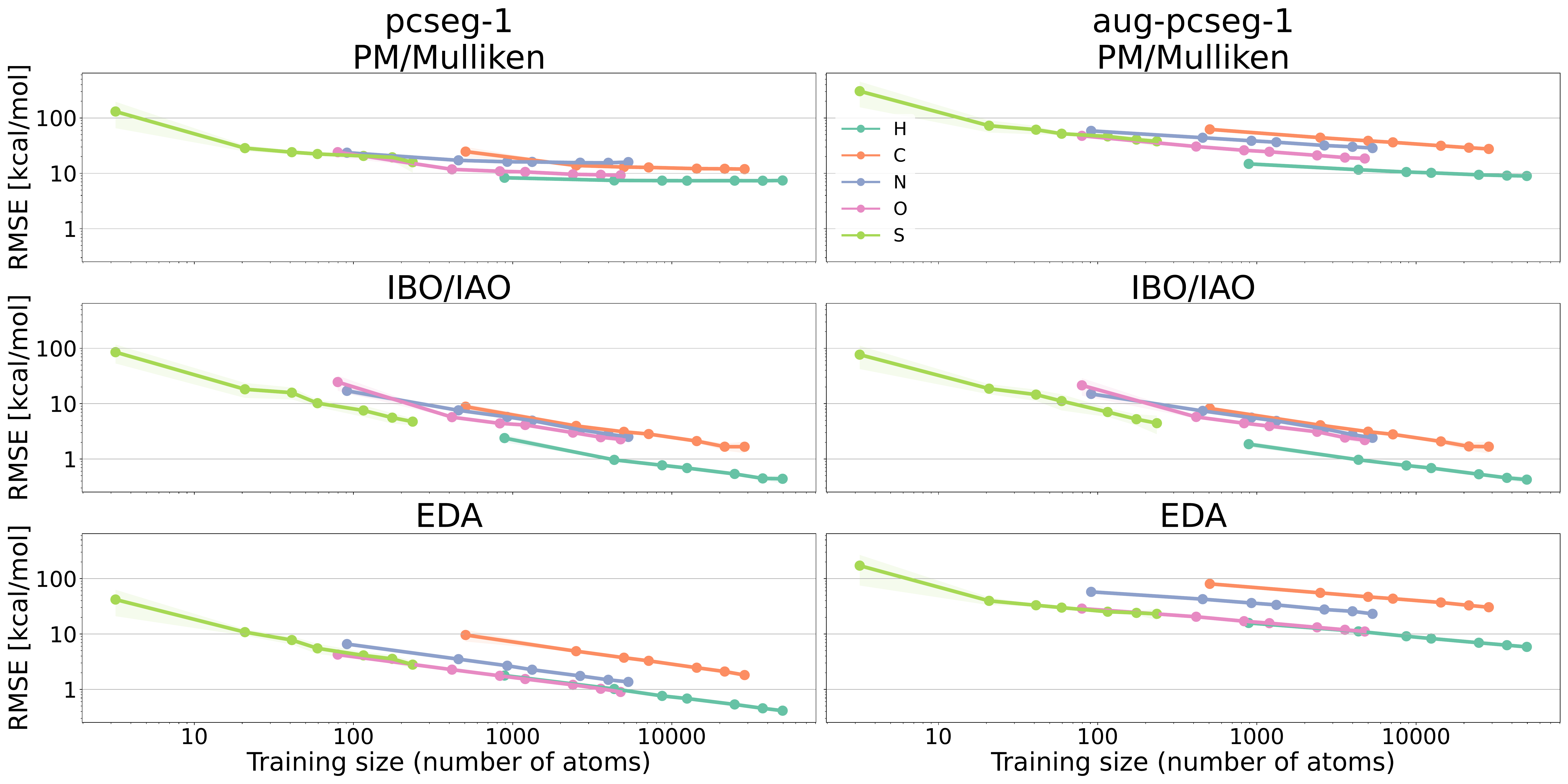}
    \caption{Training curves for each of the elements in the QM7 dataset.}
    \label{fig_4}
\end{figure}
Of much greater importance, however, is the issue of data scarcity in relation to all models trained to learn atomic energies. To illustrate the detrimental effect the heterogeneity of the different datasets for different atoms can have on model performance, Fig. \ref{fig_4} splits the total energy training curves from Fig. \ref{fig_3} into its compositional constituents. For each training curve, the combined size of the training and validation sets now depends explicitly on the number of atoms of a specific kind in these. As an example, only three sulfur atoms are on average contained in the smallest datasets consisting of 100 molecules, and the RMSEs for this element are thus enormous and of the same order across all three partitioning schemes. For the PM/Mulliken data, the training curves are all exceedingly flat, regardless of the element in question, particularly so when the size of the training data is increased, and this partitioning yields the largest overall errors at each sampling point. For hydrogen, in particular, hardly anything is learned upon offering more data to the model, although, at the same time, the element show the lowest RMSE for this partitioning. In the case of the IBO/IAO data, much steeper slopes are observed. Importantly, all training curves have approximately the same slope---and the rate of learning is the same in either of the two tested basis sets---while hydrogen once again has a significantly lower RMSE over the other elements for training data of the same size. Finally, in the pcseg-1 basis set, the EDA data reveal similar patterns as for IBO/IAO, with similar slopes for each atom. However, in this case, the training curves for all but carbon coincide, with similar RMSEs at corresponding training sizes. The prediction of carbon atomization energies, however, is far more inaccurate in comparison. Upon augmentation of the basis set, NN-based ML models are practically incapable of learning any physics on the basis of the EDA atomic contributions, cf. also the {\texttt{ANI}} version in Fig. S15 of the SI, which show much the same picture as in Fig. \ref{fig_4}.\\

We will proceed to interpret these results in the following way. For the EDA/pcseg-1 data, atomization energies for the various elements all fall within relatively narrow distributions, with the exception of carbon, for which the distribution of energies is notably broader. This is the principal reason as to why the energies of this element are more difficult for the model to learn in this basis. As for the model trained on IBO/IAO data, it learns how to predict the energies of all heavy elements at a remarkably similar rate. In this case, however, the RMSEs for hydrogen differ from the other elements, and these data are also more distinctly binned for this partitioning, cf. Fig. \ref{fig_2}. This is only true to much less extent for the PM/Mulliken counterpart, which is generally found to be unsuitable for the present purpose on account of its unsystematic decomposition of molecular energies into atomic contributions.\\

\begin{figure}[ht]
    \centering
    \includegraphics[width=\textwidth]{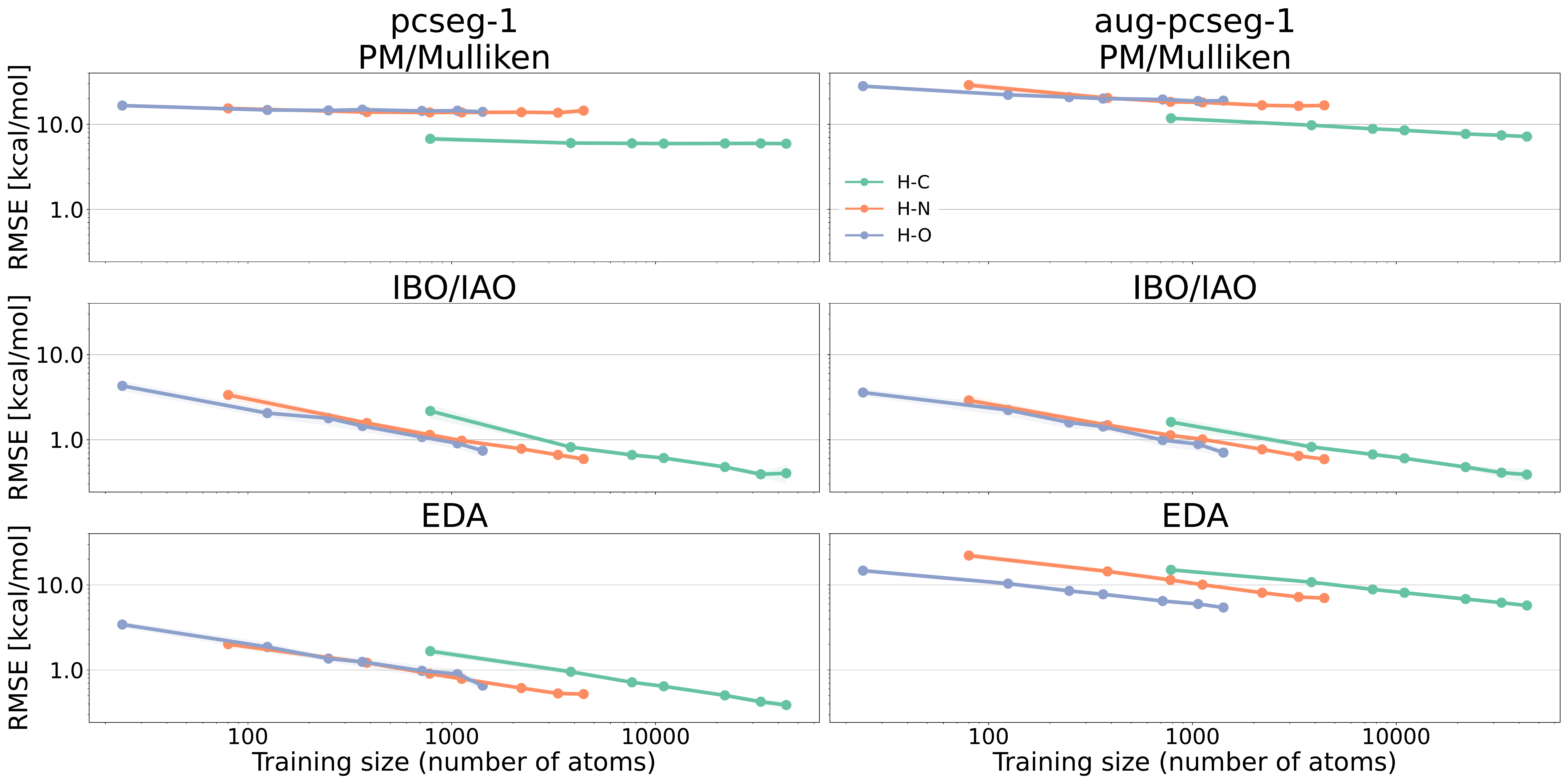}
    \caption{Training curves for the various hydrogens in Fig. \ref{fig_2}.}
    \label{fig_5}
\end{figure}
These observations may be supported by untangling the training curves in Fig. \ref{fig_3}, not on the basis of individual elements, as in Fig. \ref{fig_4}, but rather individual functional groups. As the simplest example of this, Fig. \ref{fig_5} presents training curves for the different hydrogen types in Fig. \ref{fig_2}. As is clear, a model based on PM/Mulliken data is incapable of learning the atomic energy of a hydrogen atom in either of the two basis sets, regardless of its neighbouring heavy atom. Here, the least perturbed hydrogens, that is, those bonded to carbons, have the lowest RMSEs, but variances in-between these obviously cannot be learned from the model. Due to the evident difficulties in learning these atomic energies, the NN appears to resort to focusing only on H-C energies in minimizing the loss function, as these are the most abundant. In contrast, the IBO/IAO and EDA hydrogen atomic energies can both be learned, albeit only in the unaugmented basis set in the case of the latter. Again, the training curves basically overlap for the IBO/IAO model, indicating that these are individually distinct from one another, while the curve for the H-C energies differs somewhat from the other two in the case of the EDA model in the pcseg-1 basis set. In the aug-pcseg-1 basis, on the other hand, none of the training curves overlap for the NN model trained on EDA data, rendering this partitioning scheme useless for the present purposes, on par with the PM/Mulliken scheme. The exact same conclusions may be drawn for the oxygen atoms of the most common functional groups in the QM7 dataset. In the SI, Figs. S6 and S17 present training curves for each of these (based on {\texttt{PhysNet}} and {\texttt{ANI}}, respectively), showing how the learning rates for the NN models trained on IBO/IAO data in either of the (aug-)pcseg-1 basis sets are mirror images of one another, but not so for the models trained on PM/Mulliken nor EDA data.\\

\begin{figure}[ht]
    \centering
    \includegraphics[width=\textwidth]{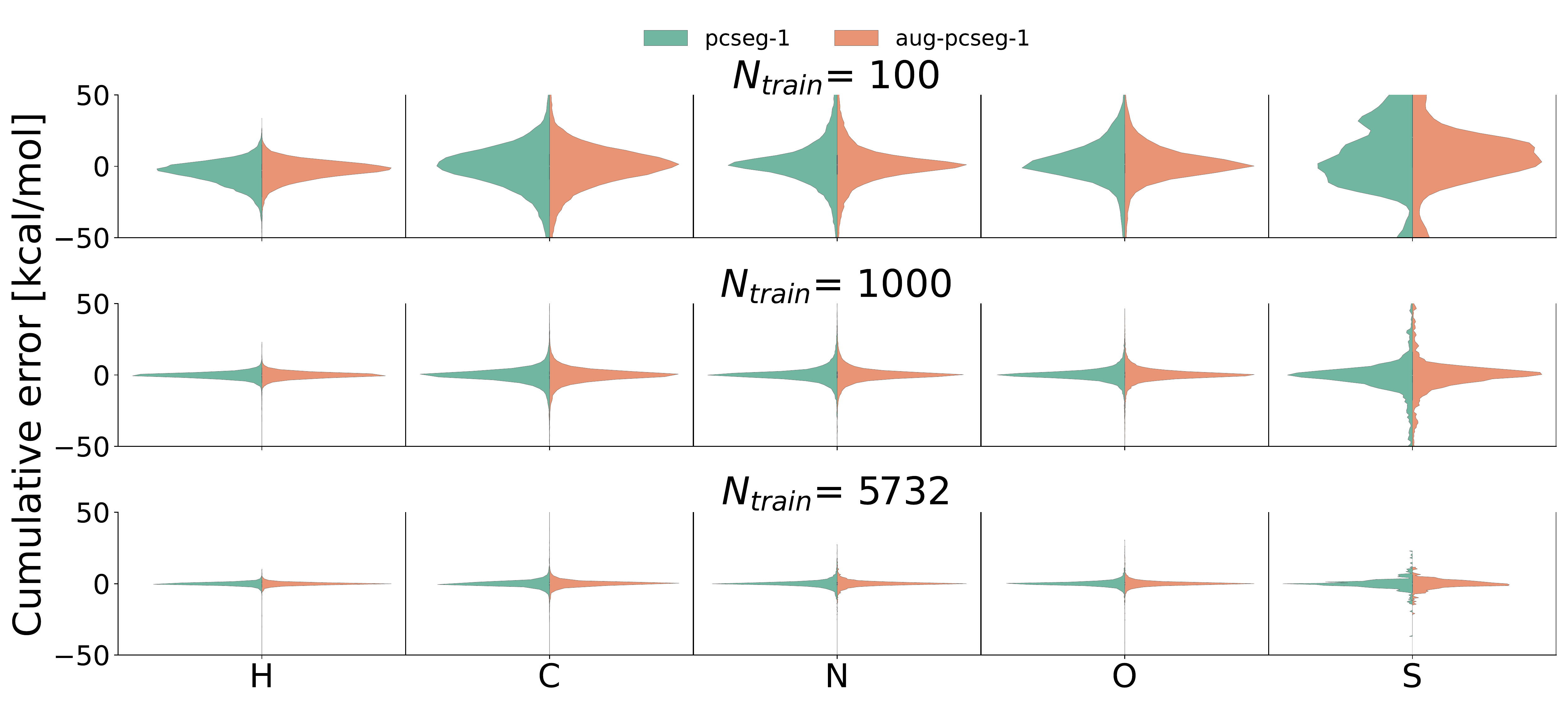}
    \caption{Cumulative contributions to molecular errors across the QM7 dataset from each of the different types of elements when training NNs on IBO/IAO atomic energies.}
    \label{fig_6}
\end{figure}
We are now in a position to probe the origin of the errors from the NN models based on atomic energies. As alluded to above in the discussion around Fig. \ref{fig_4}, the uncertainty in the sulfur predictions dominates the molecular RMSEs for low training sizes, on account of the overall scarcity of these atoms across the full QM7 dataset. In contrast, the uncertainty in the hydrogen predictions will be the lowest in this regime, but what about the accumulated errors from these atoms, given that they are the far most numerous? In Fig. \ref{fig_6}, cumulative contributions to molecular errors from each type of element are presented for three representative training sets of IBO/IAO atomization energies (PM/Mulliken and EDA versions in Figs. S7 and S8). From the results in Fig. \ref{fig_6}, we note how the S atoms indeed do make up the largest source of error, followed next by C, O, N, and H, but also that accumulation of errors may pose an issue, as evidenced, e.g., from the carbon results, at least for limited training sets. For this reason, one might speculate whether or not atoms per se identify the best possible decomposition units for the present type of partitioned machine-learned chemistry. Instead, contributions associated with unique functional moieties within molecules might exhibit a degree of physical binning on par with or even surpassing that of atomic contributions, as may energies inherent to individual chemical bonds. That being said, it should still prove instructive to closer inspect the exact manners by which different types of NN models predict local electronic structures, for instance, by comparing standard {\texttt{PhysNet}} and {\texttt{ANI}} models with the partitioned counterparts of the present work. Although atomic contributions yielded by routine NN models are often claimed to bear little to no significance on their own~\cite{isayev_ml_qc_sci_adv_2019}, some studies have made claims to the contrary~\cite{unke_meuwly_ml_nn_jcp_2018,csanyi_deringer_silicon_angew_chem_2019,mueller_montavon_nn_explanations_ieee_2022}. Figs. S9--S11 and S21--S23 of the SI compare the atomic contributions predicted by models trained on total energies with those yielded by custom models trained also on the data in Fig. \ref{fig_1}, but how exactly different ML models manage to predict finer details relating to polarization and tendencies of electrons to be shared unevenly in molecules will motivate forthcoming work from us.

\section{Summary and Conclusion}

The present study has compared a number of different ways by which to decompose the molecular energies of all members of the standard QM7 model dataset into atomic contributions. Emphasis has been placed on the emergence of trends among energies from distinctly embedded atom types, but also the stability of different schemes in their response to a change of one-electron basis set. A recent decomposition scheme of ours based on spatially localized molecular orbitals was unanimously found to expose the composition of this small chemical compound space in the most stable and physically motivated manner~\cite{eriksen_decodense_jcp_2020}, when measured in terms of its grouping and binning of discrete atomic contributions. The possibility of training tailored high-dimensional neural networks directly on atom-partitioned, rather than merely total molecular energies was further examined, and both the relevance and importance of different compositional factors for training such ML models were assessed in the process.\\

While only marginally worse in comparison with standard NN models when trained on substantial cross-sections of the full QM7 dataset, the fact that ML models built upon atomic contributions will necessarily accumulate a distribution of errors in their predictions of total molecular properties likely renders these practically unsuitable, or at least non-competitive with routine models, for this particular purpose. However, how exactly these fare in comparison with standard models for more important tasks that necessitate models to generalize well beyond (limited) training set remains to be evaluated. That being said, we still envision a number of possible practical applications for physically well-founded ML models centred around atomic contributions, such as, those based on IBO/IAO data. For instance, the availability of robust and chemically meaningful atomic contributions to total molecular energies may find use in evolutionary algorithms for the optimization of global structures, on par with previous efforts by Hammer and co-workers concerned with the spawning of molecular structures on the basis of local stability evaluations~\cite{hammer_ml_qc_jctc_2018}. On that note, genetic algorithms for the exploration of chemical compound space and their uses in the rational design of materials may likely also benefit from more local partitionings of global electronic structures~\cite{jensen_chem_space_chem_sci_2019}. While the physical insight that one may lift from models like these is valuable on its own, the potential use of structurally unique local properties as fingerprints of regioselectivity and more general reactivity proxies is also an obvious application for ML models capable of efficiently exposing the heterogeneity of local electronic structures within molecules~\cite{jensen_chem_react_chem_sci_2019,barzilay_mol_rep_jcim_2019,green_act_energy_jpcl_2020,coley_qc_aug_nn_jcp_2022}.\\

Finally, we remain highly interested in theoretical generalizations of our decomposition schemes from Ref. \citenum{eriksen_decodense_jcp_2020} that will allow for bond-wise partitionings and the machine learning of these~\cite{parkhill_intrinsic_bond_energies_jpcl_2017,persson_bondnet_chem_sci_2021}. While contributions associated with individual atoms are at risk of being intangible to most chemists, in particular, synthetic chemists who tend to make use of an altogether different language based on chemical bonds and the energies and classes of these, the present study marks an initial demonstration of the fact that physically motivated partitionings of electronic properties can indeed be efficiently learned by means of contemporary NN architectures. To that end, it remains a long-term goal of ours to further link decomposed features of chemical compound space to objects of broader interest in more practically oriented areas of chemistry. As an example, the access to contributions to molecular energies from the individual entities of a complete bond network may potentially enable partitioned models that can operate on the basis of unique functional constituents of a chemical system, e.g., in the spirit of the fragment-based ML theories of von Lilienfeld based on dictionary groupings of embedded atoms~\cite{lilienfeld_qml_jcp_2016,huang_lilienfeld_ml_amons_nat_chem_2020}. Whether simple atomic energies and the relative ranking of these in a molecular setting can prove useful for such applications, or one will ultimately need to rely on more elaborate decomposition schemes, is an area of ongoing research in our lab.

\section*{Acknowledgments}

This work was supported by a research grant (no. 37411) from VILLUM FONDEN (a part of THE VELUX FOUNDATIONS).

\section*{Data Availability}

Data in support of the findings of this study are available within the article and its SI. In addition, all raw data files and metadata have been deposited in a dedicated Zenodo repository: {\url{https://doi.org/10.5281/zenodo.7646088}}.

\section*{Supporting Information}

The supporting information (SI) collects a number of additional results in support of the present study. In Figs. S1 and S2, dependencies on one-electron basis set and DFA for all three of the tested partitioning schemes are evaluated, respectively. On par with Fig. \ref{fig_2}, Fig. S3 presents atomization energies for oxygen atoms of the most common functional groups in the QM7 dataset, while Fig. S6 presents corresponding training curves for these (as for the hydrogens in Fig. \ref{fig_5}). Figs. S4 and S5 show the sensitivity of {\texttt{PhysNet}}-based models to an augmentation of standard data by atomic energies. Figs. S7 and S8 present the same results as in Fig. \ref{fig_6}, but instead trained on PM/Mulliken and EDA atomization energies, respectively, while Figs. S9--S11 report predicted atomic energies from {\texttt{PhysNet}} NNs trained on sets of 5 seeds with either 100, 1000, or 5732 random molecules in each, respectively, using either exclusively total energies or a combination of these and data from any of the three atomic partitioning schemes as input. Finally, Figs. S12--S23 present corresponding versions of Figs. \ref{fig_3}--\ref{fig_6} and S4--S11, but based on fixed-descriptor {\texttt{ANI}} NN architectures.

\newpage

\providecommand{\latin}[1]{#1}
\makeatletter
\providecommand{\doi}
  {\begingroup\let\do\@makeother\dospecials
  \catcode`\{=1 \catcode`\}=2 \doi@aux}
\providecommand{\doi@aux}[1]{\endgroup\texttt{#1}}
\makeatother
\providecommand*\mcitethebibliography{\thebibliography}
\csname @ifundefined\endcsname{endmcitethebibliography}
  {\let\endmcitethebibliography\endthebibliography}{}


\begin{mcitethebibliography}{60}
\bibitem[Not()]{Note-1}
We used combined training and validation sets of 100, 500, 1000, 1433, 2866,
  4299, and 5732 molecules, corresponding to 1.4, 7.0, 14.0, 20.0, 40.0, 60.0,
  and 80.0 $\%$ of the total QM7 dataset, respectively.\relax
\mciteBstWouldAddEndPunctfalse
\mciteSetBstMidEndSepPunct{\mcitedefaultmidpunct}
{}{\mcitedefaultseppunct}\relax
\EndOfBibitem
\end{mcitethebibliography}
\end{document}